# Theory of Subcycle Linear Momentum Transfer in Strong-Field Tunneling Ionization

Hongcheng Ni[1,2,*], Simon Brennecke,[3] Xiang Gao,[2] Pei-Lun He,[4,†] Stefan Donsa,[2] Iva Březinová,[2]
Feng He,[4] Jian Wu,[1] Manfred Lein,[3] Xiao-Min Tong,[5] and Joachim Burgdörfer[2,‡]

[1]*State Key Laboratory of Precision Spectroscopy, East China Normal University, Shanghai 200241, China*
[2]*Institute for Theoretical Physics, Vienna University of Technology, 1040 Vienna, Austria, European Union*
[3]*Institut für Theoretische Physik, Leibniz Universität Hannover, 30167 Hannover, Germany, European Union*
[4]*Key Laboratory for Laser Plasmas (Ministry of Education) and School of Physics and Astronomy,
Collaborative Innovation Center for IFSA (CICIFSA), Shanghai Jiao Tong University, Shanghai 200240, China*
[5]*Center for Computational Sciences, University of Tsukuba, Tsukuba, Ibaraki 305-8573, Japan*

Interaction of a strong laser pulse with matter transfers not only energy but also linear momentum of the photons. Recent experimental advances have made it possible to detect the small amount of linear momentum delivered to the photoelectrons in strong-field ionization of atoms. We present numerical simulations as well as an analytical description of the subcycle phase (or time) resolved momentum transfer to an atom accessible by an attoclock protocol. We show that the light-field-induced momentum transfer is remarkably sensitive to properties of the ultrashort laser pulse such as its carrier-envelope phase and ellipticity. Moreover, we show that the subcycle-resolved linear momentum transfer can provide novel insights into the interplay between nonadiabatic and nondipole effects in strong-field ionization. This work paves the way towards the investigation of the so-far unexplored time-resolved nondipole nonadiabatic tunneling dynamics.

DOI: 10.1103/PhysRevLett.125.073202

Strong-field ionization is typically well described in the dipole approximation, in which the vector potential $A$ of the electromagnetic field as well as the electric field $F$ are assumed to be spatially homogeneous while the magnetic field vanishes. Consequently, also the Poynting vector vanishes. Thus, the strong field does not transfer linear momentum to the target. The dipole approximation is usually well justified for typical laser parameters employed in strong-field ionization in the so-called dipole oasis [1]. Remarkably, nondipole effects can be enhanced along two rather different routes—either by reducing the laser wavelength such that it approaches the atomic scale thereby probing the inhomogeneity of the electromagnetic field. This gives rise to a photoelectron angular distribution deviating from the dipole shape [2–7]. Alternatively, nondipole effects can also be enhanced by increasing the laser wavelength and/or increasing the intensity so that the motion of the liberated electron is strongly influenced by the magnetic field and radiation pressure of the laser field, resulting in linear momentum transfer along the propagation direction [8–18].

Because of the photon dispersion relation $p = \mathcal{E}/c$ ($c$: the speed of light) included in the nondipole and relativistic regime [19–21], the momentum of a single photon in the propagation direction imparted on the target, $p_z$, is very small and is usually overshadowed by the transverse momentum $p_\perp \sim \sqrt{2\mathcal{E}}$ gained by the electron from the photon energy $\mathcal{E}$, while other photon properties such as angular momentum [22–24] or helicity [25–27] have much more easily observable effects. With recent advances in detecting technologies, the small momentum shift $p_z$ in the laser propagation direction has become observable. In 2011, Smeenk et al. studied experimentally the sharing of the absorbed photon momentum between the photoelectron and the ion in tunneling ionization [8]. The photon momentum transfer could be viewed as a two-step process [15]. In the first step of tunneling, the electron and ion move together as a composite system, thus the ion gains almost all the photon momentum $I_p/c$ since it is much heavier than the electron ($I_p$: the ionization potential). In the second step of continuum motion, the liberated electron gains the momentum $E/c$, where $E$ is the final electron energy.

Further theoretical studies [28–30] found that the photon momentum transferred in the tunneling step is not given entirely to the ion due to the action of the laser magnetic field, resulting in the final electron momentum in the laser propagation direction

$$\langle p_z \rangle = E/c + \beta I_p/c, \quad (1)$$

where $\beta$ denotes the fraction of the momentum transferred to the electron during the tunneling step. Estimates for the ground-state hydrogen atom vary between $\beta = 0.3$ [28,29] and $\beta = 1/3$ [29,30].

While most previous studies focused on the momentum transfer by the entire pulse, very recently the first transverse-momentum-resolved study [15,16] and angle-resolved measurement of $\langle p_z \rangle$ [17] using an attoclock setup [31,32] became available. We present in the following the first *ab initio* quantum simulation of the subcycle linear momentum transfer in strong-field ionization resolved in time. We refer to this process as light-field-induced momentum transfer since effects of quantization of the radiation field, i.e., photonic properties, play no significant role in the strong-field regime. We show that the subcycle-resolved momentum transfer $\langle p_z(\phi_p) \rangle$ ($\phi_p$ is the attoclock angle in the polarization plane, Fig. 1) sensitively depends on the optical properties of the ultrashort pulse, most notably its carrier-envelope phase (CEP) and polarization. Conceptually importantly, it provides novel insights into the momentum sharing between the departing electron and the residual ion. Employing the backpropagation method [33–35], we are able to separate the longitudinal momentum transferred during the tunneling process and during the continuum motion of the liberated electron on a subcycle scale. Moreover, we find the time-resolved momentum transfer to be sensitive to the temporal variation of the tunneling barrier, i.e., to nonadiabatic tunneling effects. Nonadiabaticity has been known to induce energy variations across the tunneling barrier (total energy at the tunnel exit different from the initial-state energy) and shifts of the central transverse tunneling momentum from zero [34–37]. Here, we identify modulations in the linear momentum transfer, both at the tunnel exit and in the asymptotic regime, stemming from the nonadiabatic tunneling dynamics, demonstrating an interplay between the nonadiabatic and nondipole tunneling effects.

We numerically solve the time-dependent Schrödinger equation (TDSE) for strong-field ionization of helium with the single-active-electron approximation beyond the dipole approximation by including corrections to first order in $1/c$ (in atomic units). The Hamiltonian is given by [15,38]

$$H = \frac{1}{2}\left[\mathbf{p} + \mathbf{A}(t) + \frac{\mathbf{e}_z}{c}\left(\mathbf{p} \cdot \mathbf{A}(t) + \frac{1}{2}A^2(t)\right)\right]^2 + V\left(\mathbf{r} - \frac{z}{c}\mathbf{A}(t)\right), \quad (2)$$

where $(x, y)$ is the laser polarization plane (denoted by the subscript $\perp$) and $\hat{z}$ is along the direction of the laser propagation and longitudinal momentum transfer (Fig. 1), $\mathbf{p}$ is the momentum operator, $\mathbf{r}$ is the position operator, $V(\mathbf{r})$ is the effective potential for the helium atom [39], and $\mathbf{A}(t) = \mathbf{A}(t, z = 0)$ is the laser vector potential at the position of the nucleus. In the following, we choose $\mathbf{A}(t) = A_0\cos^4(\omega t/2N)[\sin(\omega t + \phi_{\text{CEP}})\mathbf{e}_x - \varepsilon \cos(\omega t + \phi_{\text{CEP}})\mathbf{e}_y]$, where $A_0$ is the peak amplitude, $\varepsilon$ is the ellipticity, $\omega$ is the central angular frequency, $N$ is the total number of cycles, and $\phi_{\text{CEP}}$ is the CEP. The corresponding electric field is defined as $\mathbf{F}(t) = -\dot{\mathbf{A}}(t)$. We use a laser pulse with a wavelength of $\lambda = 800$ nm, a total intensity of $5 \times 10^{14}$ W/cm$^2$, $N = 6$, $\varepsilon = 0.75$, and $\phi_{\text{CEP}} = 0$ or $\pi/2$, unless specified otherwise. For these laser parameters, the Keldysh parameter amounts to 0.80, corresponding to ionization in the near tunneling regime. Since a short laser pulse is used, the ponderomotive gradient from the laser focus does not transfer any net linear momentum to the electron [8]. We note that nondipole effects will increase in the midinfrared with $\lambda^2$ (see Ref. [40]). We demonstrate here that they are already observable in the near infrared. The TDSE is solved by two alternative methods to check for convergence. We use the split-operator Fourier method on a grid with 1024 points in each dimension, a grid step of 0.35 a.u., and a time step of 0.03 a.u. The simulation box is separated smoothly into an inner and an outer part by an absorbing boundary of the form $1/[1 + \exp\{(r - r_0)/d\}]$, where $r_0 = 164$ and $d = 4$ a.u. The inner part is propagated using the full Hamiltonian and the outer part using a Coulomb-free Hamiltonian. At each time step, the absorbed wave function is projected onto the nondipole Volkov solution [29,53] to incrementally obtain the momentum distribution [54,55]. Alternatively, the TDSE is also solved with the generalized pseudospectral method [56–58], where the time propagation of the nondipole contribution is treated using a Taylor expansion in combination with a split-operator method [40]. We have verified that the two numerical methods give virtually identical results. As an additional tool we employ the backpropagation method [33–35]. It allows us to extract from the full TDSE result for the flux histogram or final wave function the local characteristics including the momentum distribution at the tunnel exit. While not directly an experimental observable itself, it can provide additional key insights into the underlying tunneling dynamics. To this end, we place a sphere of 6000 evenly distributed virtual detectors [59–64] at a radius $r_{vd} = 40$ a.u. to convert the quantum flux into an ensemble of classical trajectories [40], which are subsequently propagated backward in time to retrieve the tunneling exit characteristics [26,33–35]. $r_{vd}$ is chosen such that no unphysical intercycle interference distorts the backpropagation thereby protecting the phase gradient as the local momentum [59–62]. We have checked that the results do not depend on the particular choice of $r_{vd}$ and are converged with respect to the number of virtual detectors used. The wave function is propagated for an additional cycle period after the pulse conclusion such that most ionized electronic flux reached the detector sphere.

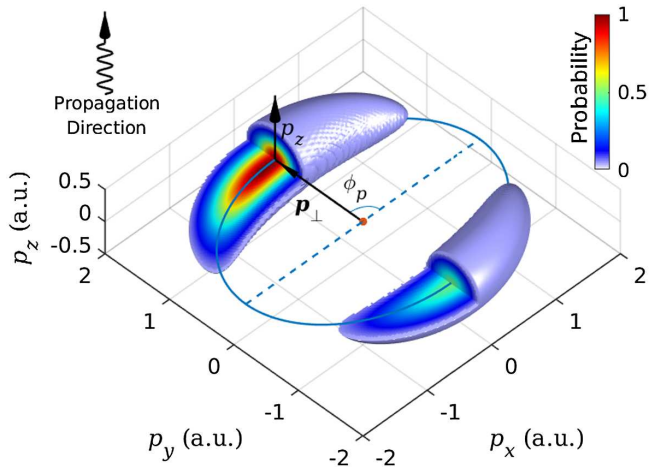

FIG. 1. Attoclock protocol for subcycle-resolved electron emission with momentum $\mathbf{p} = (\mathbf{p}_\perp, p_z)$ with the longitudinal component $p_z$ along the propagation direction and the transverse component $\mathbf{p}_\perp$ in the polarization plane. For a few-cycle pulse with ellipticity $\varepsilon$, the classical cycle-averaged radiation pressure picture suggests $\langle p_z \rangle > 0$.

In order to explore analytically the partitioning of the light-field-induced momentum transfer between electron and ion in tunneling ionization in the subcycle regime, we extend the strong-field approximation (SFA) to include both nondipole and nonadiabatic effects simultaneously (ndSFA). Accordingly, after applying the stationary-phase approximation, the tunneling-ionization rate is given by [65–67]

$$W_{ndSFA} = |\ddot{S}|^{-\alpha_Z} \exp\{2\text{Im}S\}, \quad (3)$$

with $\alpha_Z = 1 + Z/\sqrt{2I_p}$ ($Z$ is the asymptotic charge of the remaining ion) and $S = \int_{t_s}^{t_r} (\frac{1}{2}\{\boldsymbol{p} + \boldsymbol{A}(t) + (\boldsymbol{e}_z/c)[\boldsymbol{p} \cdot \boldsymbol{A}(t) + \frac{1}{2}\boldsymbol{A}^2(t)]\}^2 + I_p)dt$ [29,53] evaluated at the complex saddle-point time $t_s = t_r + it_i$, where the real part $t_r$ denotes the time the electron exits the tunneling barrier and the imaginary part $t_i$ is related to the tunneling-ionization probability. Equation (3) includes nonadiabatic as well as nondipole effects and keeps track of nonexponential prefactors. Remarkably, all of these factors leave their signature on the subcycle-resolved longitudinal momentum transfer $\langle p_z \rangle$. The present ndSFA description allows us to disentangle the longitudinal momentum transfer during tunneling from that of the free motion after tunneling with subcycle resolution.

The initial kinetic momentum at the tunnel exit $\boldsymbol{v}$ is related to the final momentum $\boldsymbol{p}$ by $\boldsymbol{v} = \boldsymbol{p} + \boldsymbol{A} + (\boldsymbol{e}_z/2c)[(\boldsymbol{p}_\perp + \boldsymbol{A})^2 - p_\perp^2] = \boldsymbol{p} + \boldsymbol{A} + (\boldsymbol{e}_z/2c)(v_\perp^2 - p_\perp^2)$. The ndSFA prediction for $\boldsymbol{v}$ can be tested against the numerical results from the backpropagation of the full TDSE solution. The time-resolved longitudinal momentum at the tunnel exit $\langle v_z(t_r) \rangle$ is found to be approximately given by [see Ref. [40], Eq. (S29)]

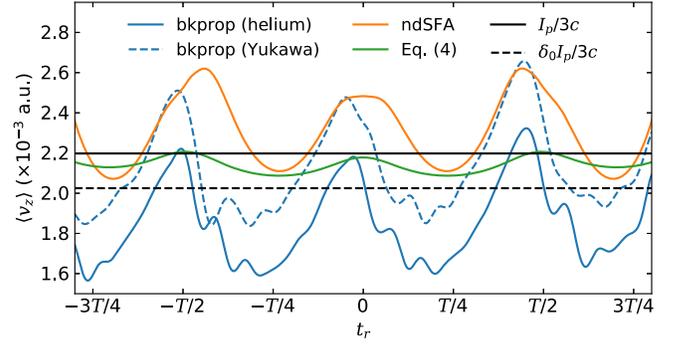

FIG. 2. Subcycle time-resolved linear momentum transfer at the tunnel exit for a sinelike pulse $\phi_{CEP} = \pi/2$. Blue curve: backpropagation of TDSE wave packet (bkprop) for the helium potential (solid curve) and the short-range Yukawa-like potential (dashed curve); orange curve: ndSFA using $\alpha_0 = 1$; green curve: analytical prediction for $\langle v_z \rangle$ [Eq. (4)] using $\alpha_0 = 1$; black solid curve: $I_p/3c$; black dashed curve: $\delta_0 I_p/3c$, with $\delta_0 = 1 - 2\alpha_0 F_0/(2I_p)^{3/2}$.

$$\langle v_z(t_r) \rangle = \frac{\tilde{I}_p(t_r)}{3c}\left[1 - \frac{2\alpha_Z F(t_r)}{(2I_p)^{3/2}}\right], \quad (4)$$

with an effective ionization potential $\tilde{I}_p(t_r) \equiv I_p + \langle v_\perp^2(t_r) \rangle/2$ accounting for the energy shift by the transverse tunneling momentum. The correction term proportional to $F$ approximately incorporates the influence from the nonexponential prefactor. The asymptotic longitudinal momentum $\langle p_z(t_r) \rangle$ follows as [16,40]

$$\langle p_z(t_r) \rangle = \Delta E/c + \langle v_z(t_r) \rangle, \quad (5)$$

providing for the subcycle decomposition of the light-field-induced longitudinal momentum transfer into one part associated with the tunneling dynamics $\langle v_z \rangle$ and another associated with the free-particle motion of the liberated electron [68] with $\Delta E = \frac{1}{2}(\langle p_\perp^2 \rangle - \langle v_\perp^2 \rangle)$. For $\langle v_\perp^2 \rangle \ll \langle p_\perp^2 \rangle$ and $\alpha_Z = 0$, Eq. (5) resembles the cycle-averaged limit [Eq. (1)]. However, significant modifications appear. Only the energy gained in the streaking field after ionization, $\Delta E$, rather than the total energy $E$, contributes to the first term [15,16]. More importantly, the longitudinal momentum transfer during tunneling $\langle v_z(t_r) \rangle$ displays a $2\omega$ subcycle oscillation and additionally a downward shift relative to the value $I_p/3c$ (Fig. 2). Scrutinizing Eq. (4), we can identify different contributions: neglecting $\langle v_\perp^2 \rangle$ and the preexponential prefactor ($\alpha_Z = 0$), Eq. (4) reduces to $\langle v_z \rangle = I_p/3c$ (black solid line). Including the prefactor for a short-range potential ($\alpha_0 = 1$), the $t_r$-independent part is shifted downward to $\langle v_z \rangle = I_p/3c[1 - 2\alpha_0 F_0/(2I_p)^{3/2}]$ (black dashed line), with the prefactor evaluated at the field peak $F_0 = A_0\omega$, where the ionization probability maximizes. It approximately agrees with the minima for the backpropagated wave packet in a Yukawa-like potential $V(r) = -[1.17822\exp(-0.2r) + \exp(-0.5r^2)]/\sqrt{r^2 + 0.14328}$ with the same $I_p$ as helium (blue dashed curve). On top of this downward shift, nonadiabaticity contributes to an upward shift $\sim \langle v_\perp^2(t_r)\rangle$ featuring a $2\omega$ subcycle oscillation reproduced by both Eq. (4) and ndSFA, where the smaller oscillation amplitude of Eq. (4) is primarily due to the expansion in $t_i$ [40]. The back-propagation from the full helium potential (blue solid curve) reveals an additional downward shift due to the presence of Coulomb attraction during the underbarrier motion of the ionizing electron.

The subcycle variation of the ionization probability as well as of the light-field-induced momentum transfer is found to strongly depend on the CEP of the pulse (Fig. 3) shown for a cosinelike pulse ($\phi_{CEP} = 0$, left column) and a sinelike pulse ($\phi_{CEP} = \pi/2$, right column). The tunneling-ionization rates obtained by the ndSFA reproduce (up to a scaling factor) quite well the rates determined by the full quantum TDSE forward propagation followed by the semiclassical backward propagation [Fig. 3(a)]. While the cosinelike pulse yields a single dominant peak at the pulse center $t = 0$ (left column), the sinelike pulse gives rise to two peaks of comparable magnitude (right column). Note that for the few-cycle elliptic pulse, the phase shift between the peaks in the $A(t)$ and the $F(t)$ fields deviates from $\pi/2$ and becomes itself time dependent [Fig. 3(a)]. Within the attoclock setting, the asymptotic momentum distribution projected onto the polarization plane provides information on the subcycle phase (or attoclock angle), $\phi_p = \arctan(p_y/p_x)$. This phase can be mapped onto the phase of the electromagnetic field through the attoclock (or streaking) principle $p_\perp \approx -A$ thereby allowing us to extract subcycle timing information from the rotating $A$ vector. The peak position of the ionization probability $P_I(\phi_p)$ is phase shifted between the full TDSE solution and the ndSFA [Fig. 3(b)] which is a well-known signature of Coulomb scattering of the outgoing electron neglected in the ndSFA. More surprisingly, the peak position of $P_I(\phi_p)$ varies strongly with $\phi_{CEP}$ [Fig. 3(b)]. In fact, this variation, virtually identical for both TDSE and ndSFA, is significantly larger than the Coulomb-field-induced shift. The momentum distribution $P(\phi_p, p_z)$ (Fig. 1) features a slight asymmetry in $p_z$ representing a signature of nondipole effects. In most previous studies [8,10,14,18,28–30], the mean shift in $p_z$ has been examined, i.e., $\langle p_z \rangle = \int d\phi_p dp_z p_z P(\phi_p, p_z)/\int d\phi_p dp_z P(\phi_p, p_z)$. We resolve the shift in $p_z$ at a given attoclock angle, i.e., $\langle p_z(\phi_p)\rangle = \int dp_z p_z P(\phi_p, p_z)/\int dp_z P(\phi_p, p_z)$ [Fig. 3(b)]. The subcycle dependence of $\langle p_z(\phi_p)\rangle$, predicted by TDSE is well reproduced by ndSFA [Eq. (3)] despite ndSFA neglects the Coulomb interaction of the outgoing electron with the residual ion. The Coulomb interaction is, thus, not important for the asymptotic longitudinal momentum transfer, because, to leading order, Coulomb-laser coupling is absent as there is no laser field in the propagation direction [16]. We note that, for linear polarization, the Coulomb effect plays a much more important role in the asymptotic longitudinal momentum transfer, both for single [9–11,14] and double ionization [18].

For ellipticities well below $\varepsilon = 1$, $\langle p_z \rangle$ features a pronounced minimum near $\phi_p \approx \pi/2$ (for details see Ref. [40]). This follows directly from Eq. (5). The energy imparted by the laser field is approximately given by $\Delta E = (\langle p_\perp^2\rangle - \langle v_\perp^2\rangle)/2 \approx A^2/2$. Since the magnitude of the electric field $F(t)$ reaches its maximum near $\phi_p \approx \pi/2$, $A(t)$ reaches a local minimum at this angle which, in turn, translates into a minimum of $\langle p_z \rangle$. The precise position of this minimum $\phi_p(p_z^{min})$ depends, however, sensitively on the CEP. While for a cosinelike pulse ($\phi_{CEP} = 0$), $\phi_p(p_z^{min})$ is located at $\phi_p = \pi/2$, for a sinelike pulse ($\phi_{CEP} = \pi/2$), the minimum is shifted to smaller $\phi_p$ [Fig. 3(b)]. Also this variation can be understood as a consequence of Eq. (5): for $\phi_{CEP} = 0$, the maximum of $F$ coincides with the minimum of $A$ [Fig. 3(a1)]; for $\phi_{CEP} = \pi/2$, the maximum of $F$ and the minimum of $A$ are slightly displaced from each other from the expected instance of time at $\omega t = \pm\pi/2$ [inset

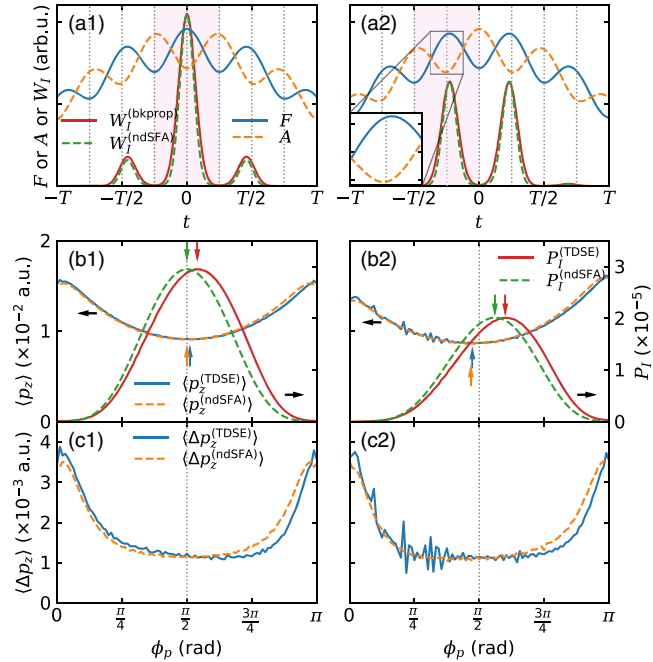

FIG. 3. Row (a): Absolute magnitude of the electric field $F(t)$ and vector potential $A(t)$ of the ultrashort few-cycle pulse with $\phi_{CEP} = 0$ (left column) and $\phi_{CEP} = \pi/2$ (right column). Also shown is the tunneling ionization rate $W_I(t = t_r)$ calculated by backpropagation (bkprop) and ndSFA using $\alpha_1 = 1 + 1/\sqrt{2I_p}$ as a function of the tunneling exit time $t_r$. Row (b): Asymptotic longitudinal momentum transfer $\langle p_z(\phi_p)\rangle$ (left ordinate) and ionization probability $P_I(\phi_p)$ (right ordinate) as a function of the attoclock angle $\phi_p$ calculated by TDSE and ndSFA for the attoclock signal corresponding to the shaded temporal half cycle in row (a). The arrows mark the phases of the maxima of $P_I$ and minima of $\langle p_z \rangle$. Row (c): Nonadiabatic longitudinal momentum transfer $\langle \Delta p_z(\phi_p)\rangle$ as a function of the attoclock angle $\phi_p$ calculated by TDSE and ndSFA.

Fig. 3(a2)] resulting in the observed phase shift. Such subcycle modulation of the linear momentum transfer will survive focal volume averaging since this timing is independent of the laser intensity [40].

Of considerable conceptual interest is now to what extent nonadiabatic tunneling effects may leave their mark on the experimentally observable asymptotic linear momentum transfer $\langle p_z \rangle$. Since $\langle p_z(t_r)\rangle = A^2(t_r)/2c - \langle v_\perp(t_r)\rangle \cdot A(t_r)/c + \langle v_z(t_r)\rangle$ [Eq. (5)], the nonadiabaticity-induced subcycle variation in $\langle v_\perp(t_r)\rangle$ not only leads to modulations of the linear momentum transfer at the tunnel exit but also is amplified in $\langle p_z \rangle$ by its coupling to the vector potential. Consequently, when subtracting from $\langle p_z \rangle$ the contributions present in the adiabatic limit, $\langle \Delta p_z \rangle = \langle p_z \rangle - A^2/2c - I_p/3c$, we find a pronounced angular modulation [Fig. 3(c)] of the residual signal that is independent of the laser CEP which should facilitate its detection [40]. Moreover, the nonadiabatic tunneling correction also leads to an overall increase in $\langle p_z \rangle$ since $\langle v_\perp(t_r)\rangle \cdot A(t_r) < 0$.

Both the angle of the minimum of $\langle p_z(\phi_p)\rangle$, $\phi_p(p_z^{\min})$, and the angle of the maximum of $P_I(\phi_p)$, $\phi_p(P_I^{\max})$, feature a systematic variation as a function of the ellipticity and the CEP [Figs. 4(a) and 4(b)]. Here, we focus on the first attoclock ionization peaks [$0 < \phi_p < \pi$, shaded area in Fig. 3(a)]. The value of $\phi_p(p_z^{\min})$ for a sinelike ($\phi_{\rm CEP} = \pi/2$) and a (−1)sinelike pulse ($\phi_{\rm CEP} = -\pi/2$) differ from each other [Fig. 4(a)]. The light-field-induced longitudinal momentum $\langle p_z(\phi_p)\rangle$ for an ultrashort pulse is a manifestly non-inversion-symmetric observable. For elliptically polarized pulses, the phase shift $\Delta\phi_p$ between the minimum of $\langle p_z \rangle$ and the maximum of $P_I$, $\Delta\phi_p = \phi_p(P_I^{\max}) - \phi_p(p_z^{\min})$, can be directly converted into a time delay $\Delta\tau = \Delta\phi_p/\omega$ [Fig. 4(c)]. We find time delays of the order of tens of attoseconds. Such time delays have indeed been observed recently, however without CEP control [17]. In view of the strong CEP dependence revealed by our simulation [Fig. 4(c)], the interpretation of this delay has remained inconclusive. In turn, future CEP-resolved measurements of $\Delta\tau$ would offer probes of the subcycle timing of the momentum transfer in unprecedented detail.

In summary, we have simulated the subcycle time-resolved light-field-induced linear momentum transfer beyond dipole approximation during strong-field tunneling ionization employing an attoclock protocol. A pronounced minimum in the momentum transfer is found along the minor axis of the polarization ellipse where the ionization probability peaks. The attoclock phase angle or timing of the minimum in the momentum transfer is shown to be strongly dependent on the carrier-envelope phase. By comparison between the backpropagation of the full TDSE solution to the tunnel exit and a novel strong-field approximation including nondipole and nonadiabatic effects, distinct contributions to the light-field-induced momentum transfer on a subcycle scale could be identified. The mean momentum shift is found to be sensitive to subexponential as well as Coulomb contributions to tunneling. Nonadiabatic corrections result in a $2\omega$ subcycle modulation of the transferred momentum at the tunnel exit and in the asymptotic region. Such nonadiabaticity-induced modulations may become directly observable in future experiments. The would represent the first direct evidence of an interplay between nonadiabatic and nondipole tunneling effects. The longitudinal momentum transfer thus promises to offer new insights into the strong-field tunneling dynamics.

This work was supported by Projects No. 11904103, No. 11774023, and No. 11574205 of the National Natural Science Foundation of China (NSFC), Projects No. M2692 and No. W1243 of the Austrian Science Fund (FWF), the Programme Quantum Dynamics in Tailored Intense Fields (QUTIF) of the Deutsche Forschungsgemeinschaft (DFG), Project No. MA14-002 of the Vienna Science and Technology Fund (WWTF), Project No. 19JC1412200 of the Shanghai Science and Technology Commission, the Fundamental Research Funds for the Central Universities, Projects No. 2018YFA0404802 and No. 2018YFA0306303 of the National Key R&D Program of China, and Project No. JP16K05495 of the Grant-in-Aid for Scientific Research of the Japan Society for the Promotion of Science. Numerical simulations were in part performed on the Vienna Scientific Cluster (VSC).

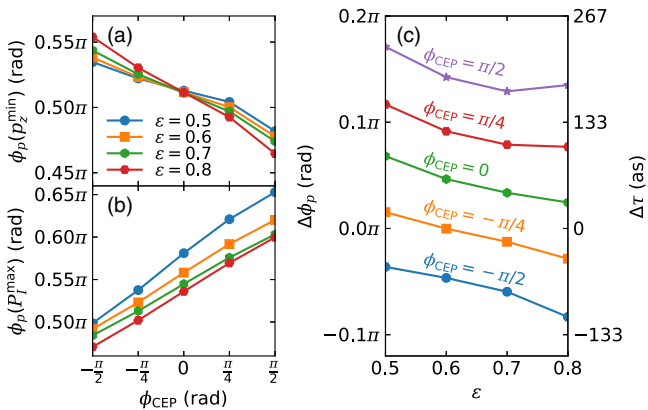

FIG. 4. The dependence of the attoclock angle $\phi_p(p_z^{\min})$ (a) and $\phi_p(P_I^{\max})$ (b) on the CEP for different ellipticities. The attoclock angle differences $\Delta\phi_p$ between $\phi_p(p_z^{\min})$ and $\phi_p(P_I^{\max})$ as a function of ellipticity $\varepsilon$ for different CEP's (c) (left ordinate) can be converted to a time delay $\Delta\tau$ in attoseconds (right ordinate).

*hcni@lps.ecnu.edu.cn
†a225633@sjtu.edu.cn
‡joachim.burgdoerfer@tuwien.ac.at

# Supplemental Material:
# Theory of Subcycle Linear Momentum Transfer in Strong-Field Tunneling Ionization

Hongcheng Ni,[1,2] Simon Brennecke,[3] Xiang Gao,[2] Pei-Lun He,[4] Stefan Donsa,[2] Iva Březinová,[2]
Feng He,[4] Jian Wu,[1] Manfred Lein,[3] Xiao-Min Tong,[5] and Joachim Burgdörfer[2]

[1]*State Key Laboratory of Precision Spectroscopy, East China Normal University, Shanghai 200241, China*
[2]*Institute for Theoretical Physics, Vienna University of Technology, 1040 Vienna, Austria, EU*
[3]*Institut für Theoretische Physik, Leibniz Universität Hannover, 30167 Hannover, Germany, EU*
[4]*Key Laboratory for Laser Plasmas (Ministry of Education) and School of Physics and Astronomy,
Collaborative Innovation Center for IFSA (CICIFSA), Shanghai Jiao Tong University, Shanghai 200240, China*
[5]*Center for Computational Sciences, University of Tsukuba, Tsukuba, Ibaraki 305-8573, Japan*

## S1. BACKPROPAGATION IN THE NONDIPOLE REGIME

In order to extract accurate information on the momentum distribution of ionized electrons at the tunnel exit, we extend the backpropagation method [1–3] to the nondipole regime. The method involves two steps: We first propagate the quantum wave packet forward in time until an ensemble of virtual detectors is reached. In the second step, semiclassical trajectories are propagated backward in time until the tunnel exit is reached. To initiate the backpropagation, the outgoing flux needs to be converted into classical trajectories when the laser field is still on. To this end, we need to find the appropriate flux associated with the Hamiltonian. The time-dependent Schrödinger equation can be written as

$$i\frac{\partial}{\partial t}\psi = \left\{\frac{1}{2}\left[\boldsymbol{p}+\boldsymbol{A}(t)+\frac{\boldsymbol{e}_z}{c}\left(\boldsymbol{p}\cdot\boldsymbol{A}(t)+\frac{1}{2}A^2(t)\right)\right]^2 + V\left(\boldsymbol{r}-\frac{z}{c}\boldsymbol{A}(t)\right)\right\}\psi, \quad \text{(S1)}$$

the conjugate of which is

$$-i\frac{\partial}{\partial t}\psi^* = \left\{\frac{1}{2}\left[-\boldsymbol{p}+\boldsymbol{A}(t)+\frac{\boldsymbol{e}_z}{c}\left(-\boldsymbol{p}\cdot\boldsymbol{A}(t)+\frac{1}{2}A^2(t)\right)\right]^2 + V\left(\boldsymbol{r}-\frac{z}{c}\boldsymbol{A}(t)\right)\right\}\psi^*. \quad \text{(S2)}$$

Multiplying Eq. (S1) by $\psi^*$ from the left, multiplying Eq. (S2) by $\psi$ from the left, and doing the subtraction, we have

$$\begin{aligned}i\frac{\partial}{\partial t}|\psi|^2 =& \frac{1}{2}\psi^*\left[\boldsymbol{p}+\boldsymbol{A}(t)+\frac{\boldsymbol{e}_z}{c}\left(\boldsymbol{p}\cdot\boldsymbol{A}(t)+\frac{1}{2}A^2(t)\right)\right]^2\psi - \frac{1}{2}\psi\left[-\boldsymbol{p}+\boldsymbol{A}(t)+\frac{\boldsymbol{e}_z}{c}\left(-\boldsymbol{p}\cdot\boldsymbol{A}(t)+\frac{1}{2}A^2(t)\right)\right]^2\psi^* \\ =& -i\boldsymbol{\nabla}\cdot\left\{\frac{1}{2}(\psi^*\boldsymbol{p}\psi-\psi\boldsymbol{p}\psi^*)+\boldsymbol{A}\left[|\psi|^2+\frac{1}{2c}(\psi^*p_z\psi-\psi p_z\psi^*)+\frac{1}{2c^2}(\psi^*\boldsymbol{p}\cdot\boldsymbol{A}\psi-\psi\boldsymbol{p}\cdot\boldsymbol{A}\psi^*)+\frac{A^2}{2c^2}|\psi|^2\right] \\ & + \frac{\boldsymbol{e}_z}{2c}(\psi^*\boldsymbol{p}\cdot\boldsymbol{A}\psi-\psi\boldsymbol{p}\cdot\boldsymbol{A}\psi^*)+\frac{\boldsymbol{e}_z}{2c}A^2|\psi|^2\right\}. \end{aligned} \quad \text{(S3)}$$

According to the equation of continuity $\frac{\partial}{\partial t}\rho + \boldsymbol{\nabla}\cdot\boldsymbol{j} = 0$, where $\rho \equiv |\psi|^2$, the probability flux $\boldsymbol{j}$ is given by

$$\begin{aligned}\boldsymbol{j} =& \frac{1}{2}(\psi^*\boldsymbol{p}\psi-\psi\boldsymbol{p}\psi^*)+\boldsymbol{A}\left[|\psi|^2+\frac{1}{2c}(\psi^*p_z\psi-\psi p_z\psi^*)+\frac{1}{2c^2}(\psi^*\boldsymbol{p}\cdot\boldsymbol{A}\psi-\psi\boldsymbol{p}\cdot\boldsymbol{A}\psi^*)+\frac{A^2}{2c^2}|\psi|^2\right] \\ & + \frac{\boldsymbol{e}_z}{2c}(\psi^*\boldsymbol{p}\cdot\boldsymbol{A}\psi-\psi\boldsymbol{p}\cdot\boldsymbol{A}\psi^*)+\frac{\boldsymbol{e}_z}{2c}A^2|\psi|^2. \end{aligned} \quad \text{(S4)}$$

If we write the wavefunction in "polar" form as $\psi = \sqrt{\rho}e^{iS}$, we find

$$\boldsymbol{j} = \rho\boldsymbol{\nabla}S + \rho\boldsymbol{A}\left\{1+\frac{1}{c}\left[\frac{\partial S}{\partial z}+\frac{1}{c}\left(\boldsymbol{\nabla}S\cdot\boldsymbol{A}+\frac{1}{2}A^2\right)\right]\right\} + \rho\frac{\boldsymbol{e}_z}{c}\left(\boldsymbol{\nabla}S\cdot\boldsymbol{A}+\frac{1}{2}A^2\right). \quad \text{(S5)}$$

Expressing the current density as $\boldsymbol{j} = \rho\boldsymbol{v}$, we obtain the local velocity $\boldsymbol{v}$ during the pulse

$$\boldsymbol{v} = \boldsymbol{\nabla}S + \boldsymbol{A}\left\{1+\frac{1}{c}\left[\frac{\partial S}{\partial z}+\frac{1}{c}\left(\boldsymbol{\nabla}S\cdot\boldsymbol{A}+\frac{1}{2}A^2\right)\right]\right\} + \frac{\boldsymbol{e}_z}{c}\left(\boldsymbol{\nabla}S\cdot\boldsymbol{A}+\frac{1}{2}A^2\right). \quad \text{(S6)}$$

Note that the Hamiltonian we use adds a "shear" to the atomic potential, effectively changing the velocity defined in this frame as compared to the lab frame. In order to obtain the observables in the lab frame, the following frame transform is required:

$$\boldsymbol{r}_{\text{lab}} = \boldsymbol{r} - z\boldsymbol{A}/c, \quad \text{(S7)}$$

$$\boldsymbol{v}_{\text{lab}} = \boldsymbol{v} + z\boldsymbol{F}/c - v_z\boldsymbol{A}/c. \quad \text{(S8)}$$

## S2. THE GENERALIZED PSEUDOSPECTRAL METHOD FOR NONDIPOLE THREE-DIMENSIONAL TDSE

The Hamiltonian used in the main text [Eq. (2)] is presented in a "sheared" gauge to facilitate TDSE simulations with the Fourier method. To check the numerical convergence of the Fourier method, we also developed the generalized pseudospectral method to solve the three-dimensional TDSE beyond the dipole approximation using a length-gauge Hamiltonian with nondipole corrections to order $1/c$:

$$H_L = \left[\frac{1}{2}\tilde{\boldsymbol{p}}^2 + V(r)\right] + \boldsymbol{r}\cdot\boldsymbol{F}(t) + \frac{z}{c}\tilde{\boldsymbol{p}}\cdot\boldsymbol{F}(t) = H_0 + H_1 + H_2, \tag{S9}$$

with $H_0 = \tilde{\boldsymbol{p}}^2/2 + V(r)$, $H_1 = \boldsymbol{r}\cdot\boldsymbol{F}(t)$, and $H_2 = \frac{z}{c}\tilde{\boldsymbol{p}}\cdot\boldsymbol{F}(t)$, where $\tilde{\boldsymbol{p}}$ is the corresponding canonical momentum. The part $H_0 + H_1$ is the Hamiltonian in dipole approximation in length gauge, and can be readily solved by the time propagator based on the second-order split-operator method [4] and the generalized pseudospectral method [5–7]. With the use of a Taylor expansion for the time propagator of the $H_2$ term, we can perform numerical simulations with the Hamiltonian $H_L$ in a similar framework. More specifically, the time propagator for $H_L$ can be expressed as

$$U(t+\Delta t) = \exp(-iH_0\Delta t/2)\exp(-iH_1\Delta t/2)\exp(-iH_2\Delta t)\exp(-iH_1\Delta t/2)\exp(-iH_0\Delta t/2). \tag{S10}$$

The propagation of $\exp(-iH_0\Delta t/2)$ is done in energy space, and $\exp(-iH_1\Delta t/2)$ is evaluated in position space. Afterwards, we continue evaluating the $\exp(-iH_2\Delta t)$ part in position space using a Taylor expansion

$$\exp(-iH_2\Delta t) \approx \sum_{n=0}^{n_{\max}} \frac{(-iH_2\Delta t)^n}{n!}. \tag{S11}$$

For the system considered in the present work, we found $n_{\max} = 8$ for each time step is adequate to obtain converged results. The momentum distribution is obtained using the same technique as in [8], except that the Volkov propagator applied to the absorbed wave packet at each time step is substituted with the nondipole version, which is calculated numerically with a similar procedure.

## S3. NONDIPOLE STRONG-FIELD APPROXIMATION

The nondipole strong-field approximation (SFA) offers the possibility to model recollision-free strong-field ionization without the computational demanding solution of the exact time-dependent Schrödinger equation. For the calculations, we use the Hamiltonian in length gauge, $H_L$ [Eq. (S9)], neglecting, however, the atomic potential in $H_0$,

$$H_L = \frac{1}{2}\tilde{\boldsymbol{p}}^2 + \boldsymbol{r}\cdot\boldsymbol{F}(t) + \frac{z}{c}\tilde{\boldsymbol{p}}\cdot\boldsymbol{F}(t). \tag{S12}$$

The initial kinetic momentum (or velocity in a.u.) at the tunnel exit $\boldsymbol{v}$ relates to the asymptotic momentum $\boldsymbol{p}$ as

$$\boldsymbol{v} = \boldsymbol{p} + \boldsymbol{A} + \frac{\boldsymbol{e}_z}{2c}[(\boldsymbol{p}+\boldsymbol{A})^2 - p^2] = \boldsymbol{p} + \boldsymbol{A} + \frac{\boldsymbol{e}_z}{2c}(v_\perp^2 - p_\perp^2). \tag{S13}$$

In the modified SFA evaluated in the saddle-point approximation, the transition rate is calculated from [9–11]

$$W_{\text{ndSFA}} = |\ddot{S}|^{-\alpha_Z}\exp\{2\text{Im}S\}, \tag{S14}$$

where $\alpha_Z = 1 + Z/\sqrt{2I_p}$ with $Z$ the asymptotic charge of the remaining ion. For a short-range potential, $\alpha_Z = \alpha_0 = 1$. In the nondipole version of SFA (ndSFA), the action including nondipole corrections to the order $1/c$ can be expressed as [12]

$$S = \int_{t_s}^{t_r}\left\{\frac{1}{2}\left[\boldsymbol{p} + \boldsymbol{A}(t) + \frac{\boldsymbol{e}_z}{c}\left(\boldsymbol{p}\cdot\boldsymbol{A}(t) + \frac{1}{2}A^2(t)\right)\right]^2 + I_p\right\}dt. \tag{S15}$$

The saddle point is given by

$$\frac{1}{2}\left[\boldsymbol{p} + \boldsymbol{A}(t_s) + \frac{\boldsymbol{e}_z}{c}\left(\boldsymbol{p}\cdot\boldsymbol{A}(t_s) + \frac{1}{2}A^2(t_s)\right)\right]^2 + I_p = 0, \tag{S16}$$

where the saddle-point time $t_s = t_r + it_i$ must be complex in order to solve Eq. (S16). After solving the saddle-point equation (S16) the photoelectron momentum distribution can be calculated directly by evaluating Eq. (S14). Reduced quantities such as the ionization probability $P_I$ as a function of the attoclock angle $\phi_p$, as shown in Fig. 3(b) of the main text, follow from integration over the remaining variables. However, in order to calculate observables as a function of the release time $t_r$ we have to integrate over all momenta that belong to this given real part of the saddle-point time. To this end, it is advantageous to perform a coordinate transformation $(p_x, p_y, p_z) \to (t_r, k_\perp, p_z)$ with the release time $t_r$ chosen as the real part of the saddle-point time $t_s$, the momentum component in the polarization plane

$$k_\perp = (\boldsymbol{p} + \operatorname{Re}\boldsymbol{A}(t_s)) \cdot (\operatorname{Im}A_y(t_s)\boldsymbol{e}_x - \operatorname{Im}A_x(t_s)\boldsymbol{e}_y) / \sqrt{(\operatorname{Im}A_x(t_s))^2 + (\operatorname{Im}A_y(t_s))^2} \tag{S17}$$

and the $z$-component of the final momentum $p_z$. The probability density in these variables is accordingly given by

$$\tilde{w}(t_r, k_\perp, p_z) = \left|\det \frac{\partial (p_x, p_y, p_z)}{\partial (t_r, k_\perp, p_z)}\right| W_{\text{ndSFA}}(\boldsymbol{p}). \tag{S18}$$

The subcycle time-resolved linear momentum transfer shown in Fig. 2 of the main text can now be obtained as the average of the initial velocity in the light propagation direction

$$\langle v_z(t_r) \rangle = \frac{\int \mathrm{d}k_\perp \mathrm{d}p_z\, v_z(t_r, k_\perp, p_z)\, \tilde{w}(t_r, k_\perp, p_z)}{\int \mathrm{d}k_\perp \mathrm{d}p_z\, \tilde{w}(t_r, k_\perp, p_z)}. \tag{S19}$$

Expansion of the vector potential $\boldsymbol{A}(t_r + it_i)$ in powers of $t_i$ [3, 13–16],

$$\boldsymbol{A}(t_r + it_i) = \boldsymbol{A}(t_r) - it_i \boldsymbol{F}(t_r) + \frac{1}{2} t_i^2 \dot{\boldsymbol{F}}(t_r) + O(t_i^3), \tag{S20}$$

allows to simplify the ndSFA and to gain additional insights. Inserting into Eq. (S16) and keeping the terms up to the second order in $t_i$ results in

$$\boldsymbol{v}(t_r) \cdot \boldsymbol{F}(t_r) = 0, \tag{S21}$$

which is the termination criterion we use for the backpropagating trajectories, and

$$t_i = \sqrt{\frac{p^2 + \left(1 + \frac{p_z}{c}\right)\left[2\boldsymbol{p}\cdot\boldsymbol{A}(t_r) + A^2(t_r)\right] + 2I_p}{\left(1 + \frac{p_z}{c}\right)\left[F^2(t_r) - \boldsymbol{v}(t_r)\cdot\dot{\boldsymbol{F}}(t_r)\right]}}. \tag{S22}$$

The ionization rate [Eq. (S14)] depends exponentially on the argument

$$\begin{aligned}\operatorname{Im}S &= -I_p t_i - \frac{1}{2}\operatorname{Re}\int_0^{t_i}\left[\boldsymbol{p} + \boldsymbol{A}(t_r + it) + \frac{\boldsymbol{e}_z}{c}\left(\boldsymbol{p}\cdot\boldsymbol{A}(t_r + it) + \frac{1}{2}A^2(t_r + it)\right)\right]^2 \mathrm{d}t \\ &\approx -\frac{\left[p^2 + \left(1 + \frac{p_z}{c}\right)\left(2\boldsymbol{p}\cdot\boldsymbol{A}(t_r) + A^2(t_r)\right) + 2I_p\right]^{3/2}}{3\sqrt{\left(1 + \frac{p_z}{c}\right)}\widetilde{F}(t_r)},\end{aligned} \tag{S23}$$

with an effective field $\widetilde{F}(t_r) = \sqrt{F^2(t_r) - \boldsymbol{v}_\perp(t_r)\cdot\dot{\boldsymbol{F}}(t_r)}$ and on the nonexponential prefactor

$$\begin{aligned}|\ddot{S}|^{-\alpha_Z} &\approx |-i(1 + p_z/c)t_i\widetilde{F}^2|^{-\alpha_Z} \\ &\approx \left[\left(1 + \frac{p_z}{c}\right)\left(v_\perp^2 + p_z^2 + 2\frac{p_z}{c}\left(\boldsymbol{p}\cdot\boldsymbol{A} + \frac{A^2}{2}\right) + 2I_p\right)\widetilde{F}^2\right]^{-\alpha_Z/2}.\end{aligned} \tag{S24}$$

Eq. (S23) can be further simplified

$$\begin{aligned}\operatorname{Im}S &\approx -\frac{\left[v_\perp^2 + p_z^2 + \frac{p_z}{c}\left(v_\perp^2 - p_\perp^2\right) + 2I_p\right]^{3/2}}{3\sqrt{\left(1 + \frac{p_z}{c}\right)}\widetilde{F}} \\ &\approx -\frac{\left\{\left(1 - \frac{p_z}{3c}\right)\left[v_\perp^2 + p_z^2 + \frac{p_z}{c}\left(v_\perp^2 - p_\perp^2\right) + 2I_p\right]\right\}^{3/2}}{3\widetilde{F}} \\ &\approx -\frac{1}{3\widetilde{F}}\left[v_\perp^2 + \left(p_z - \left(\frac{p_\perp^2 - v_\perp^2}{2c} + \frac{2I_p + v_\perp^2}{6c}\right)\right)^2 + 2I_p\right]^{3/2},\end{aligned} \tag{S25}$$

while Eq. (S24) can be accordingly approximated by

$$|\ddot{S}|^{-\alpha_Z} \approx \left[(v_\perp^2 + 2I_p)\widetilde{F}^2\right]^{-\alpha_Z/2} \left(1 + \frac{p_z}{c}\right)^{-\alpha_Z/2} \left[1 + \frac{p_z^2 + 2\frac{p_z}{c}\left(\boldsymbol{p}\cdot\boldsymbol{A} + \frac{A^2}{2}\right)}{v_\perp^2 + 2I_p}\right]^{-\alpha_Z/2}$$

$$\approx \left[(v_\perp^2 + 2I_p)\widetilde{F}^2\right]^{-\alpha_Z/2} \exp\left\{-\frac{\alpha_Z}{2}\frac{p_z}{c}\right\} \exp\left\{-\frac{\alpha_Z}{2}\frac{\left(p_z - \frac{p_\perp^2 - v_\perp^2}{2c}\right)^2}{v_\perp^2 + 2I_p}\right\}. \tag{S26}$$

Hence, we arrive at

$$W_{\text{ndSFA}} \approx \left[(v_\perp^2 + 2I_p)\widetilde{F}^2\right]^{-\alpha_Z/2} \exp\left\{-\frac{\alpha_Z}{2}\frac{p_z}{c} - \frac{\alpha_Z}{2}\frac{\left(p_z - \frac{p_\perp^2 - v_\perp^2}{2c}\right)^2}{v_\perp^2 + 2I_p} - \frac{2}{3\widetilde{F}}\left[v_\perp^2 + \left(p_z - \left(\frac{p_\perp^2 - v_\perp^2}{2c} + \frac{2I_p + v_\perp^2}{6c}\right)\right)^2 + 2I_p\right]^{3/2}\right\}$$

$$\approx \left[(v_\perp^2 + 2I_p)\widetilde{F}^2\right]^{-\alpha_Z/2} \exp\left\{-\frac{2}{3\widetilde{F}}\left[v_\perp^2 + \left(1 + \frac{\alpha_Z \widetilde{F}}{2(v_\perp^2 + 2I_p)^{3/2}}\right)\left(p_z - \left(\frac{p_\perp^2 - v_\perp^2}{2c} + \langle v_z(t_r, v_\perp)\rangle\right)\right)^2 + 2I_p\right]^{3/2}\right\}. \tag{S27}$$

The Jacobian in Eq. (S18) reads after the expansion of the vector potential in $t_i$ (S20)

$$\left|\det\frac{\partial(p_x, p_y, p_z)}{\partial(t_r, k_\perp, p_z)}\right| \approx \left|\frac{v_\perp\left[F_x(t_r)F_y'(t_r) - F_x'(t_r)F_y(t_r)\right]}{F^2(t_r)} + F(t_r)\right|. \tag{S28}$$

To this order, it does not depend on $p_z$ and, thus, has no influence in the transfer of longitudinal momentum. We therefore neglect this factor in the following.

In Eq. (S27), we introduce the partial average $\langle v_z(t_r, v_\perp)\rangle$ that is obtained by integration over $p_z$ for fixed release time $t_r$ and the perpendicular component of the velocity $v_\perp$

$$\langle v_z(t_r, v_\perp)\rangle = \frac{2I_p + v_\perp^2}{6c}\left[1 - \frac{2\alpha_Z \widetilde{F}}{(2I_p + v_\perp^2)^{3/2}}\right] \approx \frac{2I_p + v_\perp^2}{6c}\left[1 - \frac{2\alpha_Z F}{(2I_p)^{3/2}}\right], \tag{S29}$$

with the corresponding asymptotic linear momentum

$$\langle p_z(t_r, v_\perp)\rangle = \frac{p_\perp^2 - v_\perp^2}{2c} + \langle v_z(t_r, v_\perp)\rangle. \tag{S30}$$

When focusing on the temporal dynamics only, one can additionally average over the perpendicular velocity $v_\perp$ and hence obtain for a fixed release time $t_r$ the average linear momentum transfer at the tunnel exit $\langle v_z(t_r)\rangle$.

The effective field $\widetilde{F} = \sqrt{F^2 - \boldsymbol{v}_\perp \cdot \dot{\boldsymbol{F}}}$ including the time derivative of the field, $\dot{\boldsymbol{F}}$, accounts for nonadiabatic effects. Consequently, $v_\perp$ is centered at nonzero values in Eqs. (S29) and (S30). Approximating the nonadiabatic correction to the lowest order in $\dot{\boldsymbol{F}}$ in the exponent

$$W_{\text{ndSFA}} \approx \left[(v_\perp^2 + 2I_p)\widetilde{F}^2\right]^{-\alpha_Z/2} \exp\left\{-\frac{2}{3F}\left[(v_\perp^2 + 2I_p)\left(1 + \frac{\boldsymbol{v}_\perp \cdot \dot{\boldsymbol{F}}}{3F^2}\right)\right]^{3/2}\right\}, \tag{S31}$$

where the first-order term in $\boldsymbol{v}_\perp$ shifts the center of the $v_\perp$ distribution from zero to a finite value. To obtain an explicit value, we have also neglected the momentum in the laser propagation direction, which would provide only a minimal contribution to the shift in $v_\perp$. Inserting the explicit expression for the laser field and assuming a flat envelope of the pulse yields

$$\langle v_\perp(t_r)\rangle \approx \frac{\varepsilon I_p}{3A_0 \hat{F}(t_r)^3} \quad \text{with} \quad \hat{F}(t_r) = \sqrt{\cos^2(\omega t_r + \phi_{\text{CEP}}) + \varepsilon^2 \sin^2(\omega t_r + \phi_{\text{CEP}})}, \tag{S32}$$

which closely resembles the result of the Perelomov–Popov–Terent'ev (PPT) theory [17–23]. This nonadiabatic effect generates a subcycle modulation of the center of $v_\perp$ as well as $v_\perp^2$, which further results in a subcycle nonadiabatic modulation of the nondipole effect in Eq. (S31). If we approximate Eq. (S31) with a Gaussian distribution in $v_\perp$, it is easy to show that

$$\langle v_\perp^2(t_r)\rangle \approx \langle v_\perp(t_r)\rangle^2 + \frac{F(t_r)}{2\sqrt{2I_p}}. \tag{S33}$$

Distinct contributions to the linear momentum at the tunnel exit [Eq. (S29)] can be studied at different levels of approximation. If we ignore the contribution from the nonexponential prefactor ($\alpha_Z = 0$),

$$\langle v_z(t_r) \rangle \approx \frac{2I_p + \langle v_\perp^2(t_r) \rangle}{6c} \approx \frac{I_p}{3c} + \frac{\varepsilon^2 I_p^2}{54 c A_0^2 \hat{F}^6(t_r)} + \frac{F_0 \hat{F}(t_r)}{12c \sqrt{2I_p}}. \tag{S34}$$

For $\varepsilon \lesssim 1$ as in a typical attoclock setup, Eq. (S34) can be further approximated by

$$\langle v_z \rangle \approx \frac{I_p}{3c} + \frac{1}{12c} \left[ \frac{\varepsilon^2 I_p^2}{9 A_0^2} \left(1 - \varepsilon^{-6}\right) + \frac{F_0}{2\sqrt{2I_p}} (1 - \varepsilon) \right] \cos(2\omega t_r + 2\phi_{\text{CEP}}) + \frac{1}{12c} \left[ \frac{\varepsilon^2 I_p^2}{9 A_0^2} \left(1 + \varepsilon^{-6}\right) + \frac{F_0}{2\sqrt{2I_p}} (1 + \varepsilon) \right]. \tag{S35}$$

Obviously, the nonadiabaticity of strong-field tunneling ionization introduces a $2\omega$ subcycle modulation of the nondipole transfer of photon momentum at the tunnel exit, in agreement with what is found in the main text. Note that for the present laser parameters $\frac{\varepsilon^2 I_p^2}{9 A_0^2}\left(1 - \varepsilon^{-6}\right) + \frac{F_0}{2\sqrt{2I_p}}(1 - \varepsilon) < 0$, hence the modulation phase agrees with that presented in the main text as well. The temporal average of $\langle v_z(t_r) \rangle$, however, is larger than $I_p/3c$. The inclusion of the prefactor with $\alpha_Z > 0$ generates the downward shift in qualitative agreement with the results determined by the backpropagation method (black dashed line in Fig. 2 of the main text).

Nonadiabatic tunneling effects may leave their mark on the experimentally observable asymptotic linear momentum transfer $\langle p_z \rangle$. From Eq. (S30), it is easy to show

$$\langle p_z(t_r) \rangle = A^2(t_r)/2c - \langle \mathbf{v}_\perp(t_r) \rangle \cdot \mathbf{A}(t_r)/c + \langle v_z(t_r) \rangle. \tag{S36}$$

Interestingly, the nonadiabaticity-induced subcycle variation in $\langle v_\perp(t_r) \rangle$ not only leads to modulations of the linear momentum transfer at the tunnel exit but also is amplified in $\langle p_z \rangle$ by its coupling to the vector potential. Provided the linear momentum resolution suffices, a decomposition of different contributions to the angular variation shown in Fig. 3 of the main text could indeed be pursued. It may even be possible to isolate the contribution of $\langle v_z(t_r) \rangle$ by removing the contribution of $\langle \mathbf{v}_\perp(t_r) \rangle \cdot \mathbf{A}(t_r)/c$ using, e.g., a standard SFA theory. For extracting $\langle v_z(t_r) \rangle$ from $\langle p_z(\phi_p) \rangle$ (Fig. S1), we transform the angular dependence of the asymptotic linear momentum transfer $\langle p_z(\phi_p) \rangle$ to the time axis using the linear mapping between time and angle in the elliptical coordinate [24] and obtain $\langle p_z(t_r) \rangle$ for the central cycle (blue line), from which different contributions can be isolated.

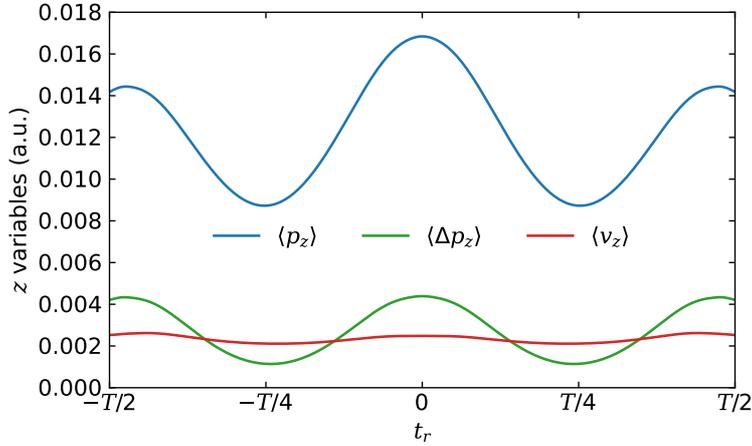

FIG. S1. Different contributions to the nonadiabatic modulation in linear momentum transfer for a sine-like pulse $\phi_{\text{CEP}} = \pi/2$ calculated by ndSFA and SFA.

## S4. ELLIPTICITY DEPENDENCE OF $\langle p_z \rangle$

The longitudinal momentum transfer $\langle p_z(\phi_p) \rangle$ features a characteristic dependence on the attoclock angle $\phi_p$. However, the exact quantity is strongly dependent on the ellipticity of the ultrashort pulse. For $\varepsilon$ well below $\varepsilon = 1$, $\langle p_z(\phi_p) \rangle$ features a pronounced minimum as discussed in the main text (Fig. S2). However, as $\varepsilon \to 1$ (circular polarization) the minimum becomes rapidly shallow and eventually turns into a shallow minimum determined by the pulse envelope.

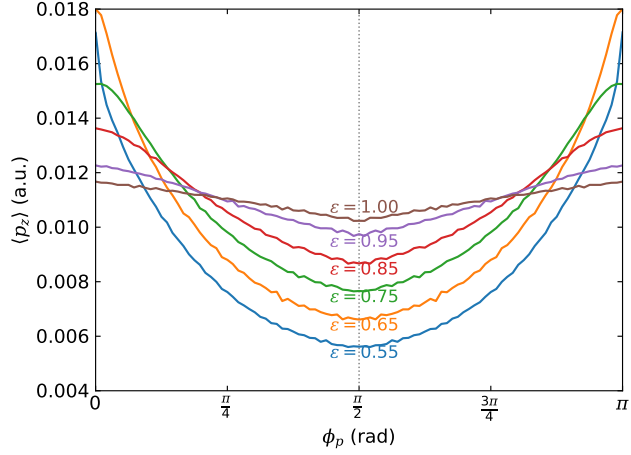

FIG. S2. Ellipticity dependence of $\langle p_z(\phi_p)\rangle$ for a cosine-like pulse $\phi_{\text{CEP}}=0$ calculated by ndSFA.

## S5. WAVELENGTH SCALING OF $\langle p_z \rangle$

In order for the nondipole transfer of linear momentum to be detected more easily, performing attoclock experiments with midinfrared laser sources is a promising route. Here, we study the wavelength dependence of the final longitudinal momentum transfer while keeping the intensity fixed, as shown in Fig. S3. Clearly, as the wavelength increases, the nondipole effect gets larger and becomes more obvious [panel (a)]. Plotting the angle-integrated final linear momentum as a function of the wavelength, we find a quadratic increase of $\langle p_z \rangle$ with the wavelength [panel (b)] and the intercept is close to but not identical to $I_p/3c$, consistent with Eq. (5) of the main text.

The ionization probability peaks near but not exactly at the minimum of the linear momentum transfer, therefore

$$\langle p_z \rangle \gtrsim \frac{I_p}{3c} + \frac{\varepsilon^2 A_0^2}{2c} = \frac{I_p}{3c} + \frac{\varepsilon^2 F_0^2}{8\pi^2 c^3}\lambda^2. \tag{S37}$$

As a result, the slope is slightly larger than $\varepsilon^2 F_0^2/8\pi^2 c^3$, which is confirmed by our results.

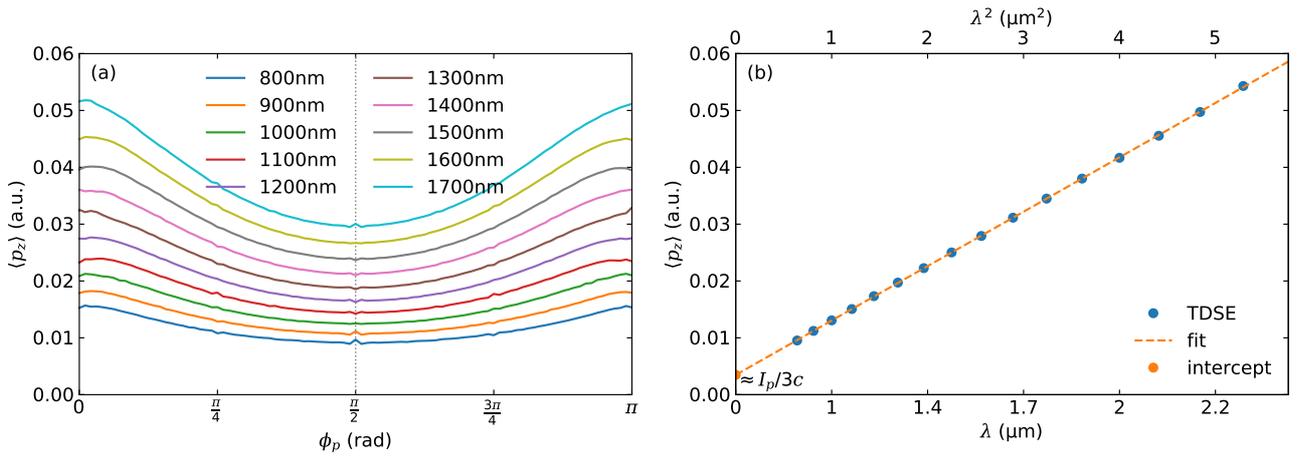

FIG. S3. Final longitudinal momentum for different laser wavelengths in (a) angle-resolved and (b) angle-integrated manner for a cosine-like pulse $\phi_{\text{CEP}}=0$ calculated by TDSE.

## S6. INTENSITY AND DURATION DEPENDENCE OF $\langle v_z \rangle$

The intensity dependence of the linear momentum transfer at the tunnel exit $\langle v_z \rangle$ is shown in Fig. S4, featuring an intensity dependence of the linear momentum transfer at the tunnel exit. The modulation depth decreases as the laser intensity increases, because the Keldysh parameter decreases, meaning the tunneling becomes more adiabatic. The most important aspect here is that the modulation is in phase for different laser intensities, which ensures that such observation survives focal volume averaging.

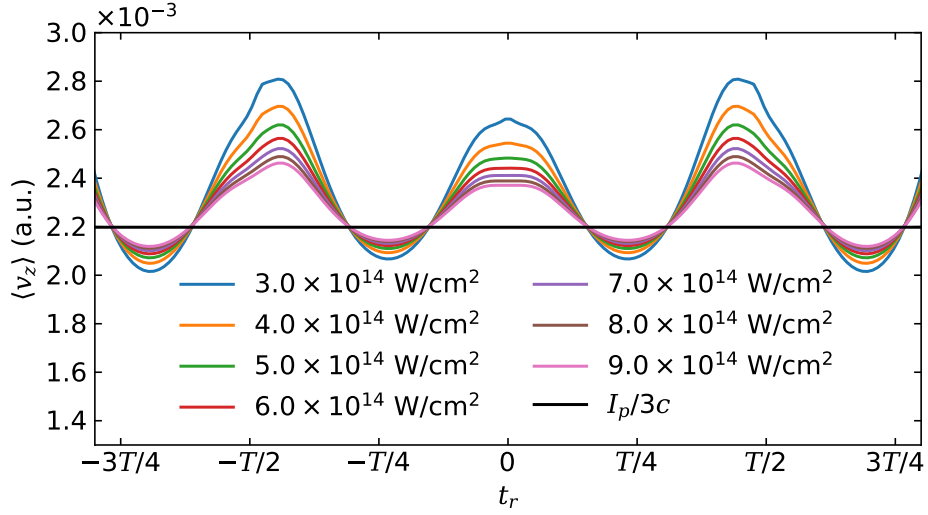

FIG. S4. Dependence of the linear momentum transfer $\langle v_z \rangle$ at the tunnel exit on the laser intensity for a sine-like pulse $\phi_{\text{CEP}} = \pi/2$ calculated by ndSFA.

The length of the pulse is found not to be a critical parameter. The variation between different laser cycles results from their relative laser intensity (or field strength) within the respective time window. When the number of cycles increases for a given peak intensity, as shown in Fig. S5, the difference between different cycles decreases. Certainly, the modulation depth is larger for short pulses at both tails due to the lower intensity and larger Keldysh parameter. However, the ionization probability in the tails is low and does not contribute much to the main ionization peaks. Therefore, the exact pulse duration is not a critical parameter for an experimental verification of the nonadiabatic nondipole effect.

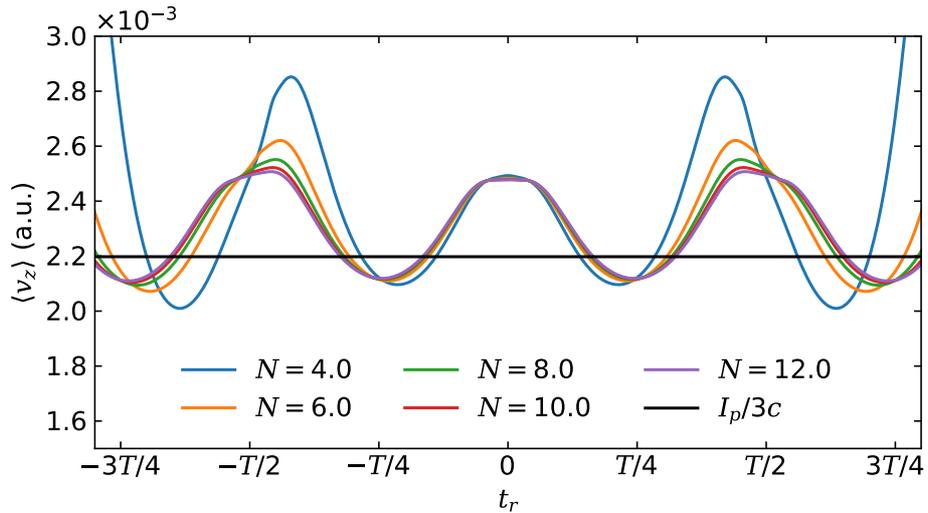

FIG. S5. Dependence of the linear momentum transfer $\langle v_z \rangle$ at the tunnel exit on the number of laser cycles for a sine-like pulse $\phi_{\text{CEP}} = \pi/2$ calculated by ndSFA.